\documentclass[%
 reprint,
 amsmath,amssymb,
 aps,
 longbibliography,
 pre,
floatfix,
]{revtex4-1}

\usepackage{graphicx}
\usepackage{dcolumn}
\usepackage{bm}
\usepackage[mathlines]{lineno}
\usepackage{cleveref}
\usepackage[titletoc,title]{appendix}

\AtBeginDocument{%
    \newwrite\bibnotes
    \def\bibnotesext{Notes.bib}
 \immediate\openout\bibnotes=\jobname\bibnotesext
  \immediate\write\bibnotes{@CONTROL{REVTEX41Control}}
    \immediate\write\bibnotes{@CONTROL{%
    apsrev41Control,author="08",editor="1",pages="1",title="0",year="1"}}
     \if@filesw
  \immediate\write\@auxout{\string\citation{apsrev41Control}}%
    \fi
}%

\begin{document}

\title{Distribution networks achieve uniform perfusion through geometric self-organization}%
\author{Tatyana Gavrilchenko$^{1,2}$}
\author{Eleni Katifori$^1$}%
\affiliation{%
$^1$Department of Physics and Astronomy, University of Pennsylvania, Philadelphia, Pennsylvania 19104\\
$^2$Center for Computational Biology, Flatiron Institute, Simons Foundation, New York, NY, 10010
}%
\date{\today}

\begin{abstract}
A generic flow distribution network typically does not deliver its load at a uniform rate across a service area, instead oversupplying regions near the nutrient source while leaving downstream regions undersupplied. In this work we demonstrate how a local adaptive rule coupling tissue growth with nutrient density results in a flow network that self-organizes to deliver nutrients uniformly. This geometric adaptive rule can be generalized and imported to mechanics-based adaptive models to address the effects of spatial gradients in nutrients or growth factors in tissues.
\end{abstract}

\maketitle

\begin{figure}
	\begin{centering}
	\includegraphics[width=0.48\textwidth]{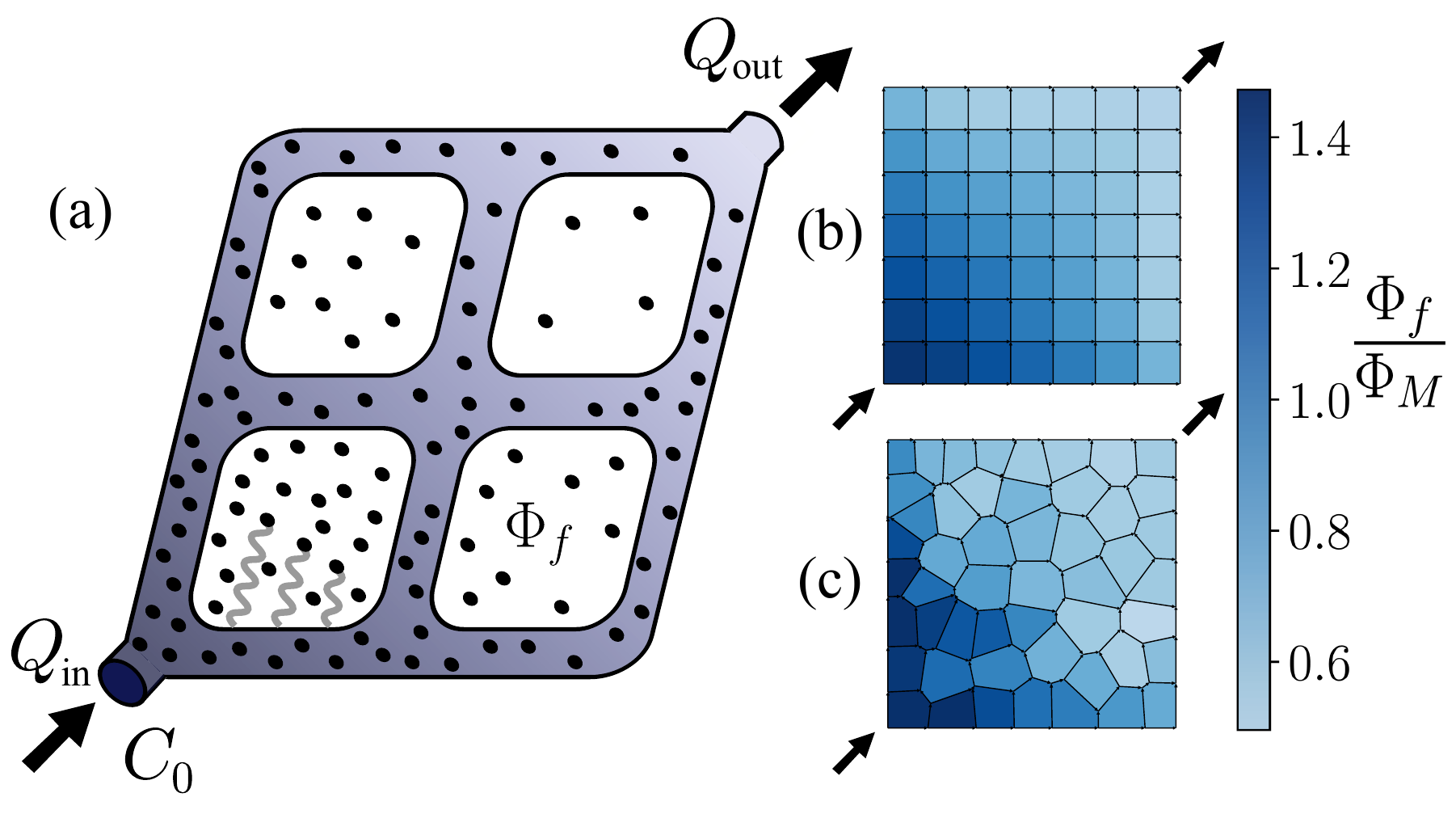}
	\caption{\textbf{Networks with a uniform topology will have a gradient in face nutrient absorption density as the nutrient concentration decays.} (a) Schematic of nutrient perfusion in a distribution network, with $\Phi_f$ signifying the nutrient absorption density per face as delivered by adjacent edges. The network model of perfusion is applied to numerically calculate $\Phi_f$ for (b) a square grid with 64 faces and (c) a Voronoi diagram with 50 faces.}
	\label{grad_demo}
    	\end{centering}
\end{figure}

Multicellular and macroscopic living organisms are continually faced with the challenge of how to uniformly distribute nutrients throughout their entire volume to maintain metabolic function while minimizing waste of resources. For this, they have evolved complex flow systems in the form of dense and space-filling networks of small vessels, termed capillaries, which distribute fluid laden with nutrients. In such perfusable systems, nutrients are carried with the flow through the capillaries and gradually diffuse across the semipermeable walls where they are used to support the metabolic needs of the tissue. In the absence of fluctuations in the flow and other mitigating factors, most network architectures, heterogeneous or uniform, will not distribute nutrients equally: in general, the tissue that is upstream will absorb more nutrients than the tissue downstream (Fig.~\ref{grad_demo}). 

When biologically-inspired microfluidic networks have been considered in the past, the emphasis has been on measuring and modeling the shear stress, resistance, flow rate, and pressure distributions, e.g as in \cite{Emerson2006, Wu2010}, and the functionality of nutrient delivery has been largely ignored. Recent work has investigated a network adaptation algorithm for uniform edge flow, showing that it is possible to tune edge conductances while maintaining the network structure to obtain equal flow over all edges \cite{Chang2019}. In the context of microvascular networks, the finite size of red blood cells may aid uniform flow \cite{Obrist2010, Chang2017, Schmid2019}. However, uniform flow does not guarantee uniform nutrient distribution: if nutrients are continually depleted by absorbing tissue, downstream tissue would generally still have a deficient supply. 

Recent work has considered the optimal architecture for uniform nutrient perfusion in plant leaves \cite{Meigel2017}. For a network with fixed edge and node positions, the authors show that adding hierarchy in edge radii yields an optimal edge conductance distribution for uniform nutrient delivery. However, this design may require an order of magnitude variation in vessel radii, and such stratification is not always possible due to developmental or physical constraints. Similarly, fabricating a network with a wide range of vessel sizes may be difficult depending on the experimental setup. For instance, the method of casting in sacrificial ink can generate networks with nonuniform edge radii, but the printed diameter is limited to a few multiples of the nozzle diameter \cite{Wu2010}.

Here we demonstrate that an arbitrary initial network can self-organize to achieve uniform nutrient perfusion using a simple geometrical and biologically plausible adaptation rule based on local information. We constrain edge radii to be equal, but allow freedom in the edge lengths and network connectivity. Similar vertex models of cell neighbor interactions are used to model the behavior of tissue sheets, typically incorporating mechanical cues for tissue adaptation such as wall tension and cell elasticity \cite{Alt2017}. These models are able to recover the geometric structure of tissue sheets \cite{Farhadifar2007, Hovcevar2009} as well as predict complex bulk properties such as collective motion of cells \cite{Bi2016, Barton2017}. In this work, we introduce an adaptive rule reminiscent of vertex models to obtain uniformly perfusing networks in an abstract setting of areas of tissue separated by channels where the extracellular fluid can flow. In this formulation, network faces do not necessarily represent individual cells, but rather, regions of tissue. Starting with an arbitrary network, the algorithm tunes vertex positions under forces that stem from differential tissue growth. A segment of tissue receiving more nutrients grows faster, effectively pushing the channels at its boundary further apart and increasing the nutrient delivery at underfed tissue. When the forces become locally balanced, the system reaches equilibrium and the final network has achieved uniform nutrient perfusion.  

A model of network perfusion must include the rate at which nutrients leave the network through the capillary membrane, or the edge absorption rate. The form of the absorption rate depends on the physical properties of the system; previous work on fungal networks has used a rate proportional to the initial edge concentration \cite{Heaton2012}. In models of oxygen transport to tissue, the basic Krogh model predicts a linear decay in oxygen concentration along a capillary \cite{Goldman2001}, and a variety of more physiologically realistic forms have been explored \cite{Salathe1978, Beard2001, Salathe2003, Goldman2006, Zimmerman2018, Erlich2019}. We stress that the equalization algorithm is applicable for any form of the nutrient concentration decay; here we follow the model of exponentially decaying concentration presented in refs. \cite{Meigel2017} and \cite{Meigel2019}. Consider a capillary with radius $r$, length $L$, and average cross-sectional flow velocity $u$. Nutrients are transported by advection along the flow and additionally by diffusion within the capillary with a diffusion constant $\kappa$. Perfusion occurs when nutrients diffuse through the capillary wall and are absorbed by the tissue; let $\nu$ be the metabolic absorption rate of the capillary membrane. Nutrient concentration decay along the capillary is shown to have the form
\begin{equation}
C(z) = C(0) e ^{-\beta z/L}
\label{edge_rule}
\end{equation}
where the decay coefficient $\beta$ is given by
\begin{equation}
\beta = \frac{24 \times \text{Pe}}{48 + \frac{\alpha^2}{S^2}} \bigg( \sqrt{1 + \frac{8S}{\text{Pe}} + \frac{\alpha^2}{6\text{Pe}S}} - 1\bigg)
\end{equation}
where Pe = $uL/\kappa$ is the Peclet number, $S = \nu L/ ru$ is the ratio of absorption rate to advection rate, and $\alpha = \nu L/\kappa $. The edge nutrient absorption rate $\phi$ is given by
\begin{equation}
\phi = \pi r^2 u C(0) \bigg( \frac{\frac{\alpha^2}{12S\text{Pe}} + 2\frac{S}{\beta}}{1 + \frac{\alpha^2}{4S\text{Pe}}}\bigg) \cdot(1-e^{-\beta})
\label{phi_full}
\end{equation}
Assumptions of the model are that the time scale of diffusion in the capillary is much shorter than the time scale of advection, $ur^2/\kappa L \ll 1$, that the capillary is long and slender, $r/L \ll 1$, and that the absorption length scale is much larger than the capillary radius, $\nu r/\kappa \ll 1$. In the limit $\beta \ll 1$, this expression simplifies to $\phi \approx 2\pi r L \nu C_0$.

We compute $\phi_{ij}$ for each edge using eq. \ref{phi_full}, first by finding the flow $Q_{ij}$ (and therefore flow velocities since $Q_{ij} = \pi r^2u_{ij}$) using current conservation at nodes and Ohm's law. Edge resistance is determined by the Hagen–Poiseuille law $R_{ij} = 8\mu L_{ij}/(\pi r^4)$, where $\mu$ is the fluid viscosity and $r$ is constant across all edges. The networks considered here have one current source and sink, and the nutrient concentration drop across each edge is computed iteratively, starting from the current source node prescribed with initial concentration $C_0$. At each node $i$, conservation of nutrient flux is obeyed with $\sum\limits_{k \textrm{, } Q_{ki}>0} C_{ki}(L_{ki}) Q_{ki} =
\sum\limits_{j \textrm{, } Q_{ij}>0} C_{ij}(0) Q_{ij}$,
where $Q_{ij}>0$ means the direction of flow is from $i$ to $j$. 

The measure of uniformity is captured by the nutrient absorption density $\Phi_f$ of each face $f$. Assuming nutrients diffuse freely within the tissue, it is defined as the nutrient received from adjacent edges scaled by face volume:
\begin{equation}
\Phi_f = \frac{1}{4 r A_f}\sum\limits_{(ij)\in f}\phi_{ij}
\end{equation}
where the thickness of the faces is set to $2r$ and a factor of $1/2$ is included because each edge supplies two faces. Let $\Phi_M$ be the metabolic demand of the tissue, a constant fixed by the cell activity levels. When $\Phi_f$ is computed for two uniform networks in Fig.~\ref{grad_demo}(b, c), a gradient is clearly present, with faces close to the source well-supplied while faces near the sink are starved. Our goal is to alter the geometry of the network in a way that eliminates the nutrient density gradient, yielding a uniformly perfusing network.

We employ a face equalization algorithm similar to a vertex model that imposes forces on the network vertices, allowing their positions to shift. A vertex experiences a repulsive force from an adjacent face with a high nutrient density and an attractive force from a low density face. Vertices on the boundary are allowed to shift, but the motion is restricted to either purely the horizontal or vertical direction to preserve the square network boundary. A vertex force becomes zero when all adjacent faces attain the same nutrient absorption density. In this way, the high nutrient density faces grow and low density faces shrink until perfusion is equalized across the network. 

We now outline the equalization procedure in detail. The force on a vertex is computed using information only from faces adjacent to the vertex. Let $\{\Phi_f\}$ be the set of nutrient absorption densities over all network faces $f$. For each vertex $i$ let $\langle \Phi_f \rangle_i\equiv \frac{1}{N_f}\sum_{f,i\in f}\Phi_f$, the mean nutrient absorption density of the $N_f$ adjacent faces. The force from face $h$ on $i$ has magnitude $\Phi_h - \langle \Phi_f \rangle_i$ and is directed away from the face centroid along the angle bisector of the face, in a fashion consistent with osmotic pressure forces in vertex models \cite{Fletcher2014}. Thus, the force points away from the face centroid if $\Phi_h > \langle \Phi_f \rangle_i$ and towards the face centroid otherwise. Vertex coordinates are shifted by a fixed step size scaled by the net force from all $N_f$ adjacent faces. If all faces adjacent to a vertex have equal nutrient density, the net vertex force is zero. After a coordinate shift, the nutrient flow and absorption through the network are recalculated, the new set $\{\Phi_f\}$ is recomputed, and the process repeats, stopping when the standard deviation of $\{\Phi_f\}$ is less than one percent of $\langle \Phi_f \rangle$, averaged over all faces. The step size is chosen to ensure that no edges overlap after vertex coordinates are shifted. If this is satisfied, we find the algorithm to be capable of achieving arbitrarily uniform networks. 

Ideally, this adaptation process is purely geometric, preserving the set of vertices and edges and changing only the vertex positions, but problems arise when network edges overlap. Edge crossings are avoided by allowing for the network topology to change via edge collapse and angle collapse, described in detail in the SM.

\begin{figure}
\begin{centering}
\includegraphics[width=0.48\textwidth]{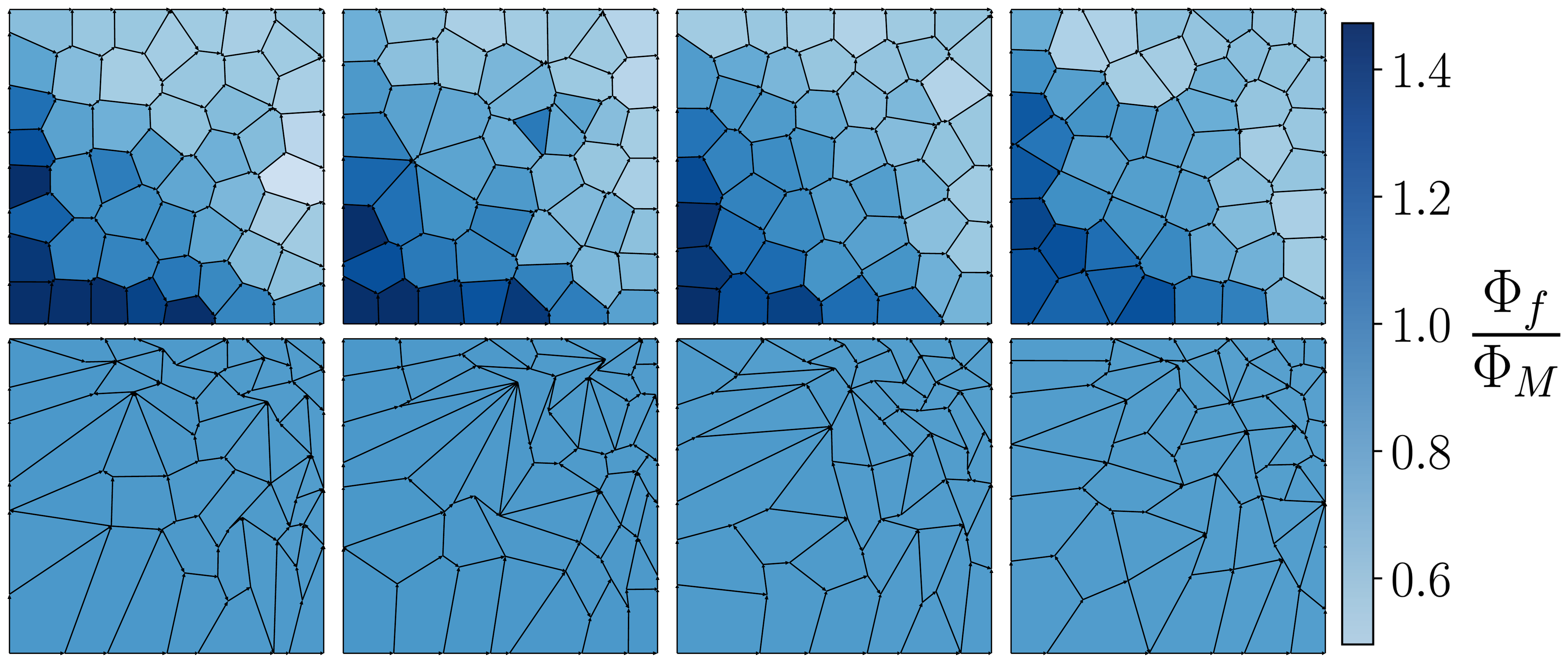}
\caption{\textbf{The face equalization algorithm results in uniformly perfusing networks.} Sample Voronoi networks with 50 faces are shown, with the corresponding equalized networks below. Equalized networks have a standard deviation in $\{\Phi_f\}$ equal to $0.01\langle \Phi_f\rangle$. A current source is located at the lower left corner and sink at the upper right corner.}
\label{f2}
\end{centering}
\end{figure}

The equalization algorithm induces a trade-off between uniformly distributed face positions and uniformly distributed nutrients. We first analyze the effects of equalization on a set of 50 Voronoi networks with 50 faces, a sample of which are shown in Fig. \ref{f2}. The side length of the full network is 10~cm, the capillary radius is $r = 0.5$~mm, and the inflow rate is $Q_{\text{in}} = 100~\mu$L~min$^{-1}$, within a realistic regime for artificial perfused vascular networks \cite{Kinstlinger2020}. We consider the transport of oxygen, which has a diffusion coefficient of $\kappa = 3\times 10^{-9}$~m$^2~$s$^{-1}$ in water and its solubility in water gives the initial concentration $C_0 = 7\times 10^{-3}$kg m$^{-3}$. The oxygen absorption rate, set by the permeability of the capillary membrane, is $\nu = 4\times 10^{-4}$~m$^{-1}$. The metabolic demand of the tissue is set to $\Phi_M = 8\times 10^{-16}$kg m$^{-3}$s$^{-1}$, with more details given in SM which includes ref. \cite{Brown2007}. We find that the initial Voronoi networks have $\Phi_f$ decaying away from the source but a uniform distribution of face centroids. During equalization, face positions are nonuniformly shifted towards the sink, as discussed in SM. The equalization algorithm acts like a growth induced pressure, as vertex forces arise from differences in $\Phi_f$ in adjacent faces.

\begin{figure}[htp]
\begin{centering}
\includegraphics[width=0.47\textwidth]{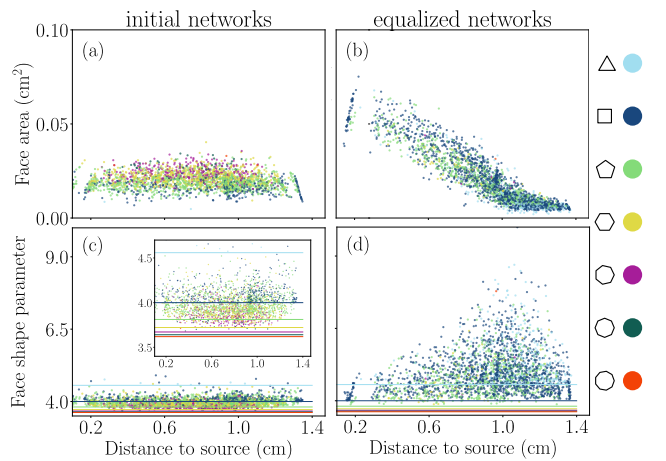}
\caption{\textbf{Equalized networks have smaller and elongated faces near the sink}. (a, b) Face area as a function of Euclidean distance from the face centroid to the source at $(0,0)$. Color indicates face shape, classifying polygonal faces by their number of sides. Stratification by color in the initial networks indicates that polygonal faces with a larger number of sides tend to be larger than faces with fewer sides. (c, d) Face shape parameter as a function of distance to the source. The shape parameters of equilateral polygons are shown as solid horizontal lines.}
\label{f3}
\end{centering}
\end{figure}

We find that equalized networks attain morphological features that allow for uniform perfusion, namely an asymmetric distribution of the face sizes and shapes. First, uniform networks have larger faces near the source and smaller faces near the sink. Fig.~\ref{f3}~(a, b), shows the distribution of face area along the space of the network. In the initial networks, face area is strongly correlated with the polygon type, i.e. the number of sides of the face, but not with face location. This contrasts with the uniformly perfusing networks, where there is no clear correlation between face area and polygon type, but there is a strong correlation between area and location, with large faces near the source and small faces near the sink. After the equalization process, edges that provide more nutrients feed larger faces, and the nutrient-rich edges are all near the source, causing large faces in that region. Nutrients in the network unavoidably decay from the source to the sink, but an asymmetric distribution of face sizes allows for uniformly perfusing networks.

The relation between the face location and elongation is identified in terms of the face shape parameter. This is a measure of the compactness or elongation of a polygon, and has been used to classify planar tilings and predict the jamming behavior of tissues \cite{Hovcevar2009, Bi2016, Siber2018}. Defined as the dimensionless quantity ${p_0=\text{perimeter}/\sqrt{\text{area}}}$, it is minimal for regular polygons. For example, $p_0=2(\sqrt[4]{3})^3 \approx 4.56$ for an equilateral triangle and $p_0=4$ for a square, and a non-regular triangle or quadrilateral will have strictly larger $p_0$.

The change in the face shape parameter distribution between the initial and equalized networks is shown in Fig.~\ref{f3}(c, d). For the initial networks, $p_0$ has no correlation with the location in the network, but is dictated by the number of sides in the face. Moreover, all faces are nearly regular, since the shape parameter for each type of polygonal face lies close to the minimal shape parameter for that regular polygon. For the uniformly perfusing networks, there is no longer a strong dependence between the face $p_0$ and polygon type, but there is a correlation between the shape parameter and location: faces near the source tend to be more compact and faces by the sink tend to be more elongated. Since faces far from the source have less nutrient absorption per edge, they compensate to supply the same nutrient absorption density by increasing the total face perimeter per area. 

\begin{figure}
    \centering
    \includegraphics[width = \linewidth]{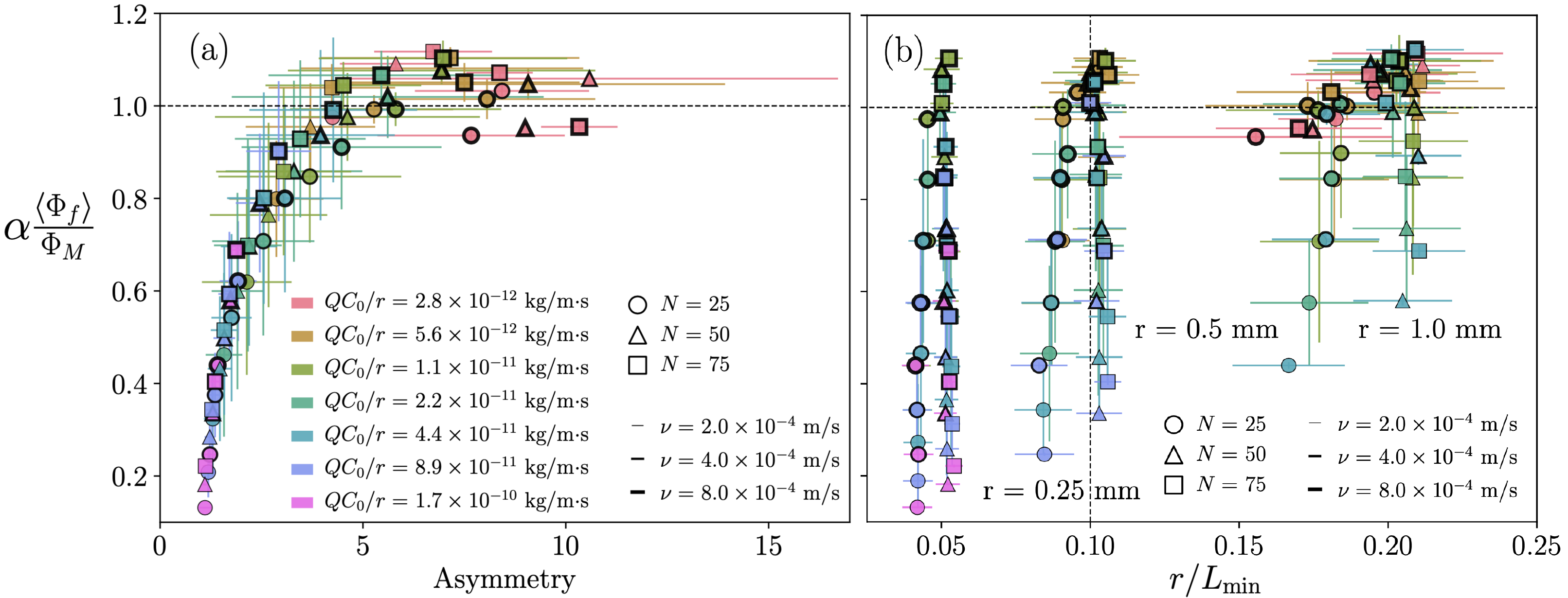}
    \caption{\textbf{Plotting the network asymmetry and efficiency reveals a geometric scaling law}. $\langle\Phi_f \rangle /\Phi_M$ is normalized by the dimensionless parameter $\alpha = \frac{r}{r^*}\frac{Q^*}{Q}\frac{C_0^*}{C_0}$, where $r^* = 0.5$~mm, $Q^* = 100~\mu$L~min$^{-1}$, and $C_0^* = 7\times 10^{-3}$kg m$^{-3}$. We consider three different values of $r, C_0$,  and $Q$; see S2 for details. Each data point is the average over 10 networks equalized to three percent uniformity with error bars marking one standard deviation.}
    \label{f4}
\end{figure}

We have shown that uniform perfusion can be achieved in general regardless of the initial network architecture, but in practice the resulting nutrient field must also meet the metabolic demands of the tissue. We consider steady state nutrient perfusion, assuming that once the nutrients leave the capillaries and enter the tissue they are able to diffuse freely. We denote the quantity $\langle \Phi_f \rangle/\Phi_M$ the network efficiency: an equalized network meets the metabolic demands of the tissue if the efficiency is greater than one. To evaluate the equalized networks, we propose a measure of network asymmetry, defined by the ratio of the number faces closer to the outlet to the number of faces closer to the inlet. The asymmetry of the initial Voronoi networks is about one since the faces are evenly distributed in space (see Fig. \ref{f3}a), and is larger for equalized networks, as the distribution of face size and compactness is skewed; for reference, networks presented in Fig. \ref{f2} have a mean asymmetry of $3.0$. This single measure serves to bundle the distributions of face size and compactness. High asymmetry is an indicator that a high number of topological transitions have occurred during equalization, which means that the initial network was not suitable for the choice of parameters, as we discuss further on. We find that the equalized network efficiency has a clear relation with the asymmetry, and that renormalizing appropriately, the data collapse. The network efficiency obeys the linear scaling relation $\langle \Phi_f \rangle/\Phi_M \propto Q C_0/r$ (Fig.~\ref{f4}a), indicating the geometric nature of the equalization process.

Finally we discuss how to select a suitable initial network before implementing the equalization procedure for an experimental perfusion network. As previously stated, a final equalized network is suitable for perfusion if $\langle \Phi_f \rangle/\Phi_M > 1$. While theoretically there is no upper limit on the asymmetry of the final network, large deformations to the initial structure can be problematic. As the asymmetry of a network increases, it is more likely to have small faces, and therefore short edge, by the network sink. The model expression for $\phi$ becomes invalid once the capillaries become too short, i.e. when $r/L \ll 1$ fails to be true. This sets a limit on the geometry of a suitable network. We set $L_{\text{min}}$ to be the average of the top ten percent of shortest network edges, and consider the length condition to be met if $r/L_{\text{min}} < 0.1$. These two conditions are met in the upper left quadrant in Fig.~\ref{f4}b, narrowing the section of parameter space that can be used to generate networks with an adequate nutrient supply. While the parameters $\nu$ and $\kappa$ depend on the material properties of the tissue and capillary walls and are thus difficult to tune, $r$, $Q_{\text{in}}$, $C_0$, and $N$ can easily be modulated in the experimental setup. Fig.~\ref{f4}b show that the capillary radius is the dominant factor for selecting an initial network; if $r > 0.5$~mm the finalized network is likely to have edges that fail the length criterion. Increasing the number of faces tends to increase the efficiency but also $r/L_{\text{min}}$, therefore there is a balance to be struck in choosing the density of the initial network. Finally, if $r$ and $N$ are suitable, increasing $Q$ of $C_0$ will increase the efficiency while maintaining the network geometry. The selected region of parameter space can reasonably be attained experimentally.

In light of recent advances in artificial microvascular devices \cite{Miller2012, Kolesky2016, Wang2018, Redd2019}, it is necessary to develop a theory of network design for uniform perfusion over an extended space. We have presented a self-organizing method for generating uniformly perfusing networks from arbitrary initial networks using a geometric local adaptation rule. The equalization algorithm successfully achieves uniformity on all networks we have tried; the networks presented have a standard deviation in ${\Phi_f}$ that is $1\%$ of the mean of ${\Phi_f}$. During the equalization process, this model is free to explore the space of configurations, utilizing network geometry as a degree of freedom. The equalization algorithm works by tuning the global distribution of network faces, forming large, compact faces by the source and small elongated faces by the sink. The network efficiency scales linearly with the capillary radius, the nutrient inflow rate, and the initial nutrient concentration. Finally, a suitable initial network must be chosen for the equalized network to fulfill metabolic demands.

The algorithm presented here is a proof of principle that a network can harness local geometric responses to stimuli to achieve a state of uniform perfusion. While a global property like total energy dissipation is difficult for the network to monitor, biological networks are often able to measure select local edge properties. For instance, capillary networks can sense flow velocities though the wall shear stress and internal nitric oxide, produced in response to low oxygen levels, and modulate vessel diameters to ensure that adequate operational levels are maintained \cite{Hu2012, Meigel2019, Jensen2009}. Network growth models that use feedback from tissue oxygenation levels have been shown to mimic features seen in natural vasculature networks \cite{Secomb2013}. Since our equalization algorithm requires information only from the nearest neighboring faces to compute vertex forces, it is a local computation. We do not expect natural networks to use this exact mechanism, but because this local algorithm consistently results in uniformity it may be an example from a larger class of natural adaptation strategies for resource distribution.

\medskip
TG would like to thank T. Machon for invaluable comments on the manuscript and I. S. Kinstlinger for discussions on fabrication of experimental systems. This research was supported by the NSF Award PHY-1554887, the University of Pennsylvania Materials Research Science and Engineering Center (MRSEC) through Award DMR-1720530, the University of Pennsylvania CEMB through Award CMMI-1548571, and the Simons Foundation through Award 568888. EK would like to acknowledge the Burroughs Wellcome Fund for their support.

\bibliography{capillaries}
\end{document}


\title{Distribution networks achieve uniform perfusion through geometric self-organization: Supplementary Material}
\author{Tatyana Gavrilchenko$^{1,2}$}
\author{Eleni Katifori$^1$}%
\affiliation{%
$^1$Department of Physics and Astronomy, University of Pennsylvania, Philadelphia, Pennsylvania 19104\\
$^2$Center for Computational Biology, Flatiron Institute, Simons Foundation, New York, NY, 10010
}%

\maketitle

\onecolumngrid

\section*{S1: A\lowercase{llowed topological rearrangements in the equalization algorithm}}

\begin{figure}[ht]
    \centering
    \includegraphics[width=0.95\textwidth]{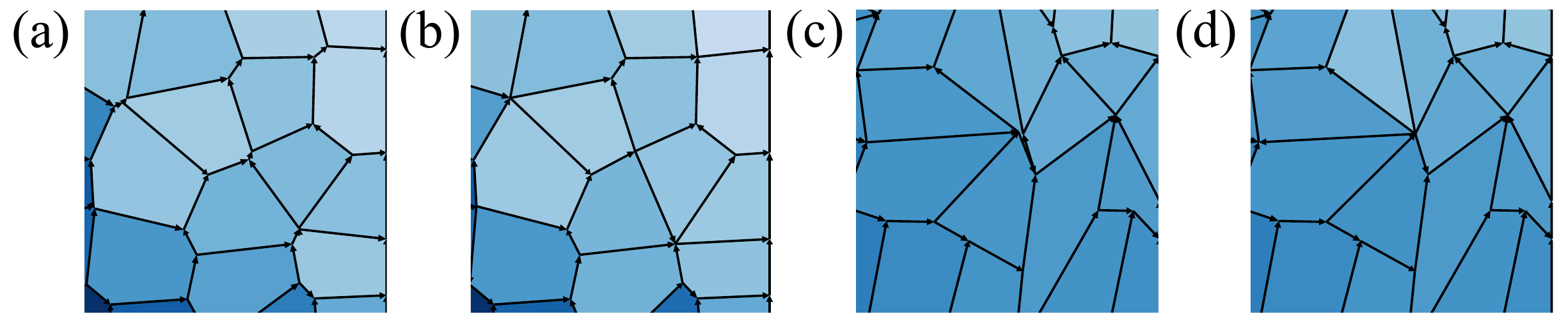}
    \caption{Examples of the two types of topological changes allowed by the equalization algorithm. An edge beneath the length threshold $l = 1/30 L_N$ is shown in the center of (a) and is shrunk to a vertex in (b). An angle measuring below the five degree threshold is shown in (c) and is collapsed to a final configuration shown in (d).}
    \label{rearrangement_ex}
\end{figure}

The equalization algorithm attempts to preserve the network topology, but allows two types of topological changes in order to maintain planarity. First, if an edge becomes shorter than a threshold length (set to $1/30$ of the full network side length) then the edge is removed and the two vertices connected by the edge are merged. Second, if an angle formed by vertices $abc$ becomes smaller than a threshold angle (set to 5 degrees), then the longer edge $ab$ is removed and a new edge $ac$ is added, effectively collapsing the acute angle. Edge collapse and angle collapse are analogous to topological T2 and T1 transitions, respectively. An example case for using each of the rearrangements is shown in Fig.~\ref{rearrangement_ex}. 

\section*{S2: D\lowercase{iscussion of parameter choices}}

For parameter selection, we chose to stay within the range used for experimental perfusion setups, for example like those presented in \cite{Kinstlinger2020}. In this system, hollow channels are carved into a cell-laden agarose gel. An oxygenated broth flows through the network through a single input and output point, and oxygen is able to diffuse into the gel. Networks are three-dimensional and attain a range of sizes, with the largest having dimensions 33 mm x 15 mm x 14 mm. Channel diameter can range from 300 to 900 $\mu$m. For the simulations, the capillary radius is set to 0.5 mm and the network length and width are set to 10 cm, and the network height is set to 1 mm, which results in a similar volume of living tissue as in the experimental system. The experimental flow rate must be such that the shear stress on the capillary walls does not destroy the material. Experimentally, a value of 100 ml min$^{-1}$ was used, which is 1.6 mm$^3$/s. The value $Q = 0.8$ mm$^3$/s was chosen as the benchmark value for simulations.

The metabolic demand of the tissue depends on the density of cells and the metabolic demand per cell, which varies depending on the type of cell used. As an estimate, metabolic consumption rate is computed from the oxygen consumption rate given by \cite{Brown2007} for HepG2 liver cells with a cell number density of $10^6$ mL$^{-1}$.

For Fig. 4 in the manuscript, the diffusivity of oxygen is fixed at $\kappa = 3\cdot 10^{-4}$  m s$^{-1}$ and the rest of the parameters are varied as shown in Table \ref{T1}. 

\begin{center}
\begin{table}
\begin{tabular}{p{1cm}p{2cm}p{2cm}p{2cm}p{2cm}} 
r & mm & 0.25 & 0.5 & 1.0 \\
$Q$ & m$^3$/s & 0.8 & 1.6 & 3.2 \\
$C_0$ & kg/m$^3$ & $3.5\cdot 10^{-13}$ & $7\cdot 10^{-12}$ & $1.4\cdot 10^{-11}$ \\
$\nu$ & 1/m & $2 \cdot 10^{4}$ & $4\cdot 10^{4}$ & $8\cdot 10^{4}$ \\
$N$ & & 25 & 50 & 75
\end{tabular}
\caption{Parameter name, units, and the three chosen values used to generate Fig. 4 in the main text. The intermediate value for each of $r$, $Q$, and $C_0$ is chosen as the benchmark value to use for normalization.}
\label{T1}
\end{table}
\end{center}


\section*{S3: E\lowercase{equalization algorithm nonuniformly shifts vertices towards the network sink}}

\begin{figure}
\begin{centering}
\includegraphics[width=0.80\textwidth]{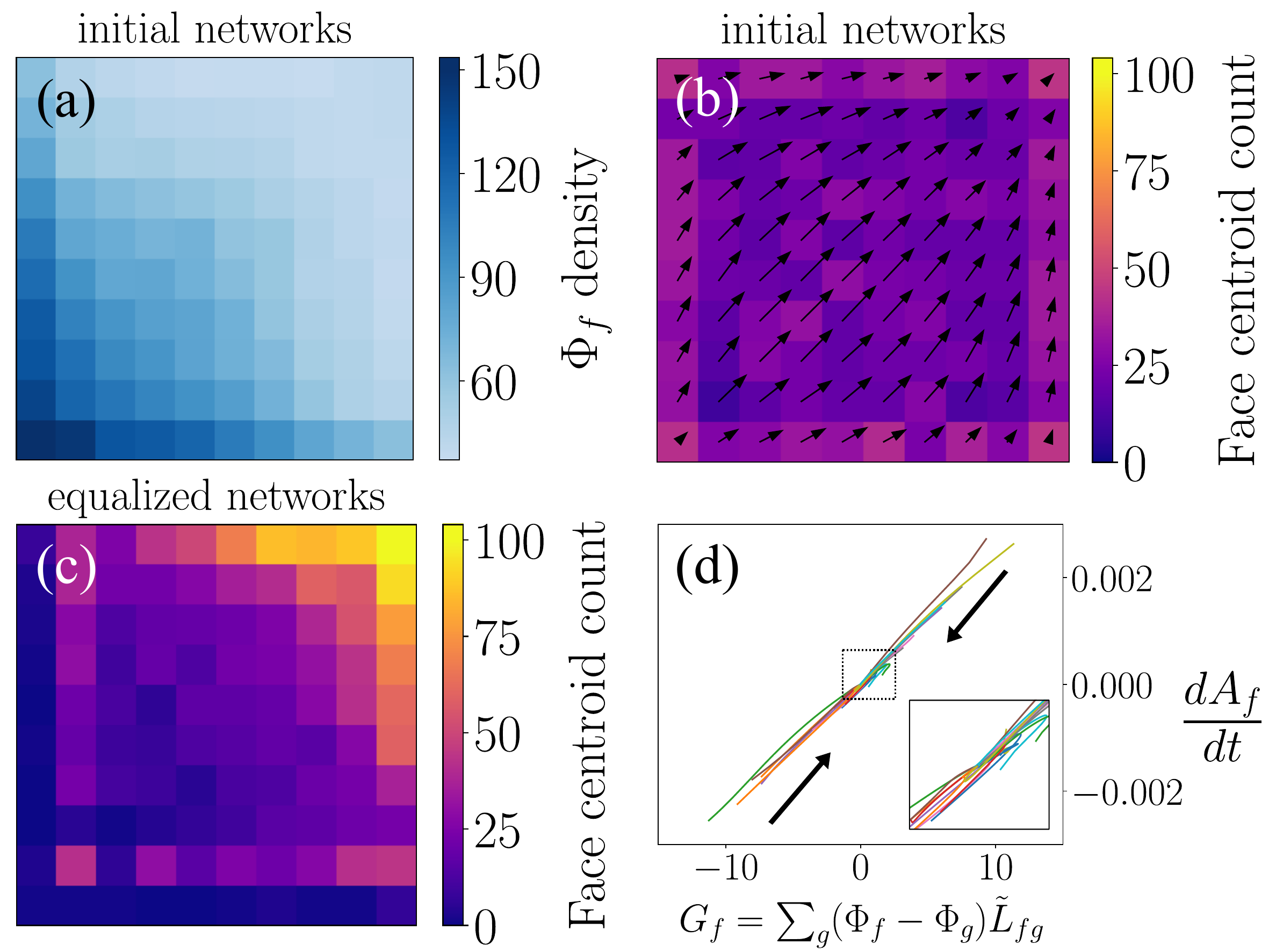}
\caption{\textbf{Face centroids shift towards the sink as the face absorption density is equalized.} Information from 50 networks is spatially binned, showing trends across the initial and equalized network ensembles. (a) The initial Voronoi networks have the density $\Phi_f$ decaying away from the source, but (b) a uniform distribution of face centroids. Arrows indicate the average movement of the face centroids during the equalization procedure, with arrow length proportional to the mean displacement, showing a general motion of faces towards the sink. (c) Equalized networks have a higher density of faces by the sink. (d) $\Phi_f$ serves the role of pressure in the usual vertex model, as shown by the trajectories of face areas over the course of the equalization for a single network. Faces grow or shrink faster in the early stages, but as networks are equalized the change in area for all faces approaches zero.}
\label{f3}
\end{centering}
\end{figure}

To analyze the effects of equalization, we use a set of 50 Voronoi networks with 50 faces. The set is equalized to produce an ensemble of uniformly perfusing networks. Figure~\ref{f3}(a)-(c) outlines the general motion of network faces during the equalization algorithm, ultimately showing the trade-off between uniformly distributed face centroid positions and uniformly distributed nutrients. These figures show the cumulative value of face centroid position and face $\Phi_f$ over all networks, binned in a $10\times 10$ grid indicating space. In the initial networks, nutrients are concentrated by the source (Fig.~3(a)) and the face centroid distribution is approximately uniform (Fig.~3(b)). The motion of the face centroids during equalization is shown, with arrows indicating the mean direction of motion for a single spatial bin. The equalized networks have a higher density of face centroids by the nutrient sink, shown in Fig.~\ref{f3}(c), but a uniform distribution of $\Phi_f$ density (not shown). Thus, equalization under the local adaptive rule serves to exchange uniformity in face centroids for uniformity in face nutrient absorption density by shifting the network face centroids towards the sink.

Finally, Fig.~\ref{f3}(d) shows that the equalization algorithm acts like a growth induced pressure, as vertex forces arise from nutrient density differences in adjacent faces. Let $G_f$ represent a quantity akin to the cumulative force a face is subjected to: $G_f = \sum_g(\Phi_f-\Phi_g)\tilde{L}_{fg}$, where $\tilde{L}_{fg}$ is the length of the edge separating faces $f$ and $g$. We find a nearly linear relation between $G_f$ for a face $f$ and its change in area over time, $dA_f/dt$. Growing faces start with $dA_f/dt > 0$ and shrinking faces start with $dA_f/dt < 0$, and as the network is equalized the change in area approaches zero for all faces.

\section*{S4: T\lowercase{he equalization algorithm is effective for a variety of initial network choices}}

In the main text, the initial network ensemble under consideration is a collection of Voronoi diagrams with 50 points. Here we demonstrate that the equalization algorithm is effective for other types of initial networks, namely regular tilings, alternative Voronoi-like networks with an added spring force drawing algorithm, and percolated irregular triangular tilings. 

\begin{figure}[ht]
    \begin{centering}
    \includegraphics[width= 0.9\textwidth]{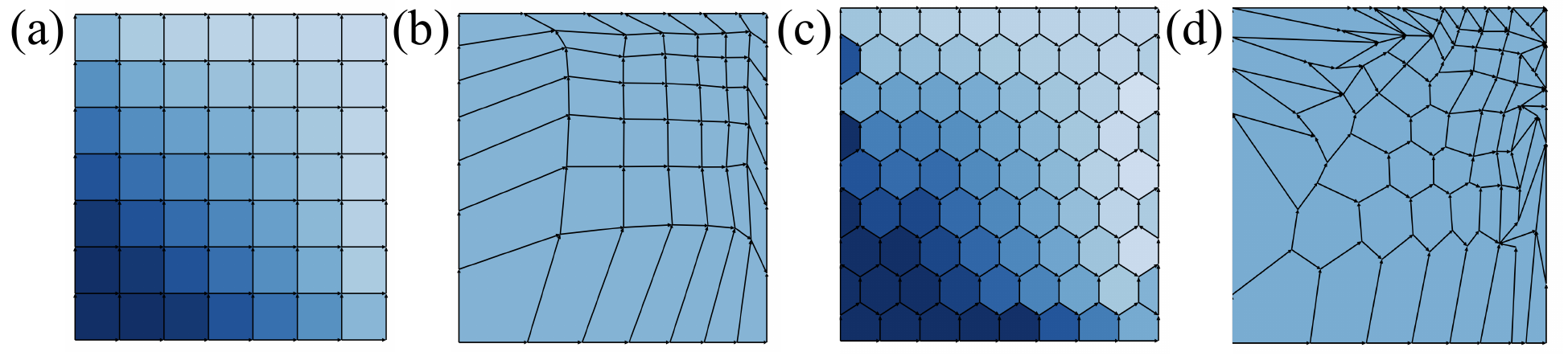}
    \caption{The equalization algorithm implemented on regular square and regular hexagonal tiling. For the square tiling, the algorithm preserves topology, while for the hexagonal tiling, the algorithm uses the two types of allowed topological rearrangements, the faces in the equalized network are no longer only hexagonal.}
    \end{centering}
\end{figure}

\begin{figure}[ht]
    \centering
    \includegraphics[height=0.32\textwidth]{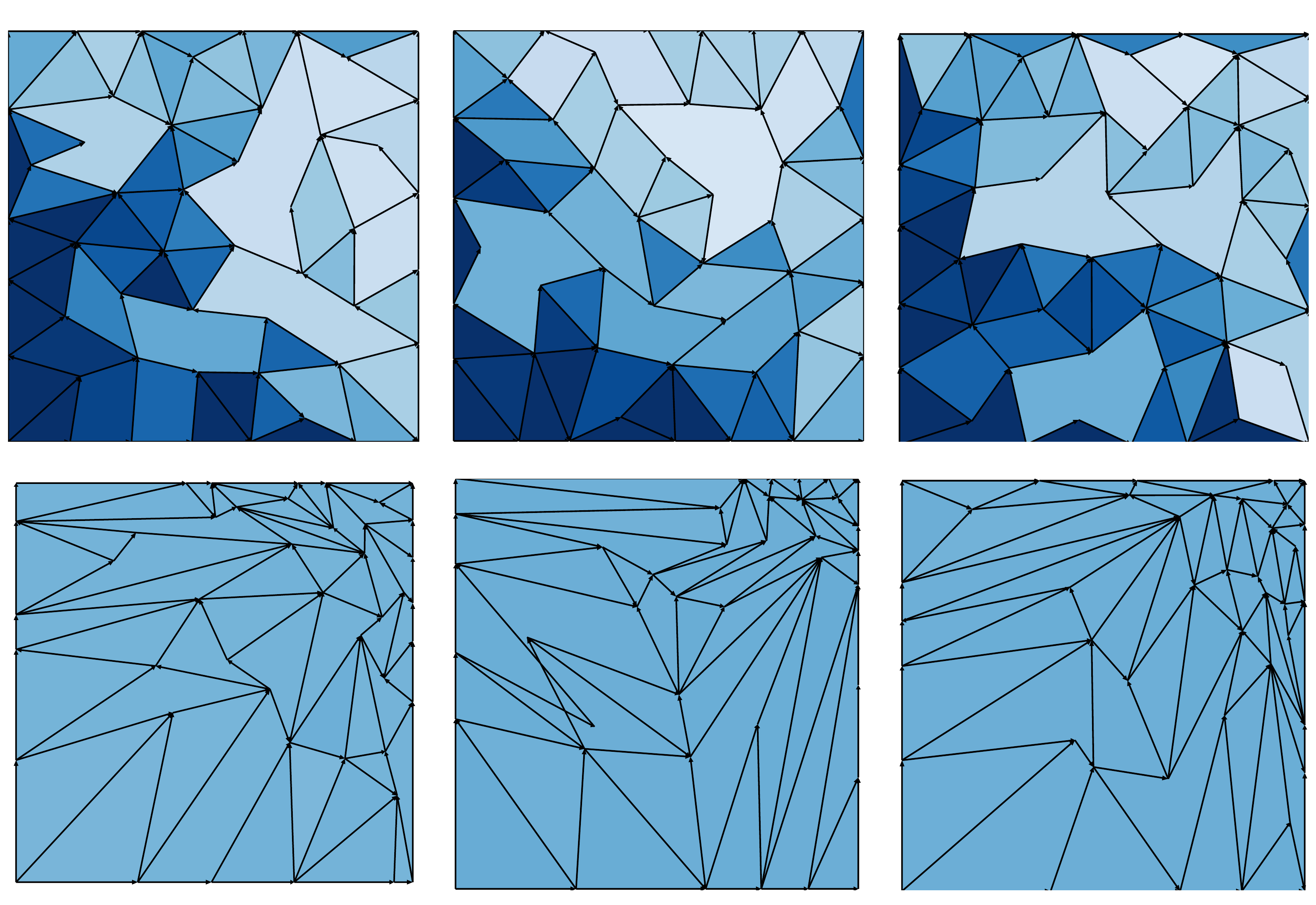}
    \includegraphics[height=0.32\textwidth]{fig_s4_b.pdf}
    \caption{Figs. 2 and 4 from the main text recreated for an initial ensemble of polygonal networks created by taking the dual network of an irregular triangular tiling with an added spring force drawing algorithm to create more regular polygons.} 
\end{figure}

\begin{figure}[ht]
    \centering
    \includegraphics[height=0.32\textwidth]{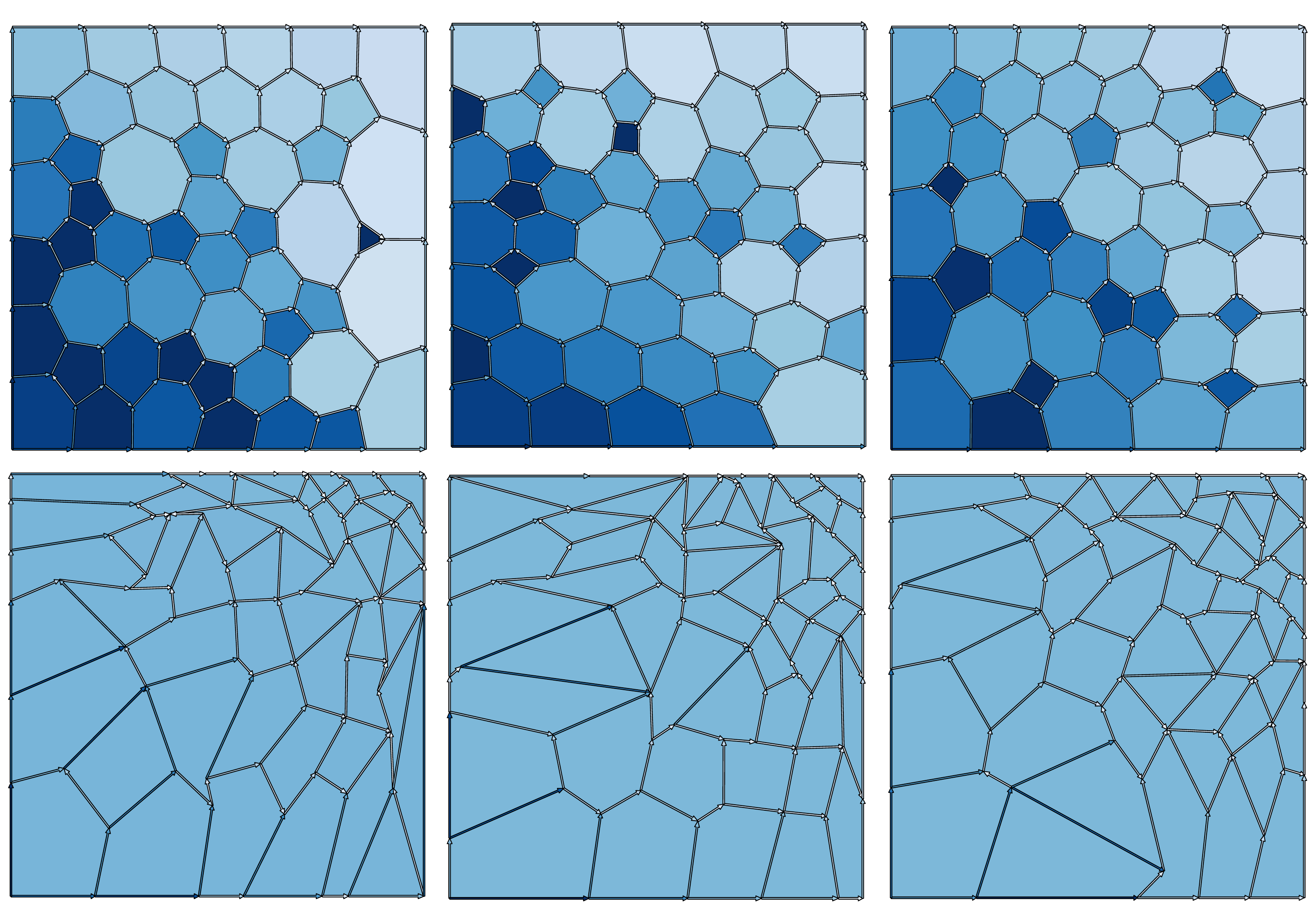}
    \includegraphics[height=0.32\textwidth]{fig_s5_b.pdf}
    \caption{Figs. 2 and 4 from the main text recreated for an initial ensemble of percolated irregular triangular tilings. The irregular triangular tilings are created from the triangularization of an initial randomly-placed set of 50 points on the unit square, which are spaced more uniformly by imposing a repulsive potential on a point from all other points. Then 25 edges from the bulk are removed at random, resulting in networks with 30 to 40 faces. }
\end{figure}

\FloatBarrier

\bibliography{capillaries}